\documentclass[twocolumn]{openjournal}
\usepackage{orcidlink}
\usepackage{ulem}
\usepackage{lineno}
\usepackage{threeparttable}
\usepackage{comment}
\usepackage{color}
\usepackage{hyperref}
\hypersetup{
	colorlinks=true,
	urlcolor=blue,    
	linkcolor=black,  
	citecolor=blue,   
}

\newcommand{\orcidauthor}[3]{\author{\href{http://orcid.org/#1}{#2$^{#3}$}}}
\begin{document} 
\interfootnotelinepenalty=10000

\title{Detached Circumstellar Matter as an Explanation for Slowly-Rising Interacting Type Ibc Supernovae\vspace{-1.5cm}}


\orcidauthor{0000-0002-8215-5019}{Yuki Takei}{1,2,3,*}
\orcidauthor{0000-0002-6347-3089}{Daichi Tsuna}{4,2}

\affiliation{$^{1}$Yukawa Institute for Theoretical Physics, Kyoto University, Kitashirakawa-Oiwake-cho, Sakyo-ku, Kyoto, Kyoto 606-8502, Japan}
\affiliation{$^{2}$Research Center for the Early Universe, Graduate School of Science, University of Tokyo, Bunkyo-ku, Tokyo 113-0033, Japan}
\affiliation{$^{3}$Astrophysical Big Bang Laboratory, RIKEN, 2-1 Hirosawa, Wako, Saitama 351-0198, Japan}
\affiliation{$^{4}$TAPIR, Mailcode 350-17, California Institute of Technology, Pasadena, CA 91125, USA}

\thanks{$^*$E-mail: \href{mailto:yuki.takei@yukawa.kyoto-u.ac.jp}{yuki.takei@yukawa.kyoto-u.ac.jp}}


\begin{abstract}
Some hydrogen-poor (Type Ibc) supernovae (SNe) are known to have massive circumstellar matter (CSM) that are well detached from the star.
Using the open-source code \texttt{CHIPS}, we construct a grid of models of SN Ibc interacting with detached CSM, inspired by a recently proposed scenario of such CSM generated by a mass eruption and feedback process from the leftover star.
We find that interaction with detached CSM can produce a slowly rising phase in the light curve seen in some interacting SN Ibc, which is difficult to reproduce by interaction with continuously distributed CSM down to the star.
We also show that SNe having double peaks in their light curves with timescales of months (e.g., SN 2022xxf) can be explained by radioactive decay of $^{56}$Ni/$^{56}$Co, followed by interaction with detached CSM. 
\end{abstract}
\maketitle

\section{Introduction}
\label{sec:introduction}
Some massive stars are known to undergo episodic mass loss just before core-collapse to form a dense circumstellar matter (CSM) \citep[e.g.,][]{Chugai04,Pastorello07,Mauerhan13,Moriya14,Bruch21,Strotjohann21}. Supernovae (SNe) interacting with dense CSM show bright light curves and prominent narrow emission lines, powered by collision of the SN ejecta with the CSM. Studying the CSM is essential for understanding massive star evolution, as it provides direct insights into the dramatic pathways leading to its death.

For hydrogen-poor stripped envelope SNe, the structure of the dense CSM is found to be diverse. The representatives of such interacting stripped-envelope SNe has been Type Ibn \citep[e.g.,][]{Pastorello08}, and the more recently discovered Type Icn \citep[e.g.,][]{GalYam22}. The signs of ejecta-CSM interaction from the very early phase of the SN, such as narrow lines of helium/carbon/oxygen and the typically fast light curve rise of a few days, indicate the existence of dense CSM close to the star ($\lesssim 10^{15}$\,cm, for a typical ejecta velocity of $\lesssim 0.1c$).

More recently a class of double-peaked SNe with a luminous second peak in the optical has been identified \citep[e.g.,][]{Kuncarayakti22,Kuncarayakti23,Das23,Chen_2024,Sharma_2024}. In the broad-lined Type Ic SN~2022xxf \citep{Kuncarayakti23} representative of this class, the second peak slowly emerged a month after the first peak presumably powered by the radioactive decay of $^{56}$Ni in the ejecta. Some of these SNe show signatures of CSM interaction around the second peak, implying a massive detached CSM far from the progenitor ($\gtrsim 10^{15}$--$10^{16}\,{\rm cm}$) in contrast to typical Type Ibn/Icn SNe. There are also some SNe Ib which show re-brightening likely due to interaction with detached hydrogen-rich CSM, such as SN~2014C and SN~2019yvr \citep{Milisavljevic15,Margutti17,Ferrari24}.

While these recent events have presented strong evidence for the existence of detached CSM in some SNe, detailed simulations on the diversity of the light curves expected from these CSM are yet to be conducted. In this work we simulate a grid of bolometric light curves from detached CSM of various radii, using an open-source code \texttt{CHIPS} \citep{Takei22,Takei24}.
Using a model for the density profile of the detached CSM inspired from the model of \citet{Tsuna_Takei_2023}, we show that interaction with detached CSM can lead to distinctive light curve features, such as slow rises and long-duration double-peaked structures. Such structures are generally difficult to reproduce by typically assumed density profiles of continuous CSM down to the star.

This work is constructed as follows. In Section \ref{sec:model}, we present our model for constructing the detached CSM and calculating light curves powered by CSM interaction, and investigate their basic properties. In Section \ref{sec:Observation} we compare our model to observations of interacting SN Ibc with slowly-rising light curve features, namely a Type Ibn SN~2010al and the aforementioned SN~2022xxf. Section \ref{sec:Discussion} provides discussion, including some of the limitations of our model that can be improved in future works.

\section{Our Model}
\label{sec:model}
Here we present our framework for constructing the detached CSM and modeling the bolometric light curves.
We first summarize our light curve modeling code \texttt{CHIPS} for interacting SN Ibc \citep{Takei24}, and the main characteristics of the light curves obtained in that work.
The model by \citet{Takei24} considered CSM that are continuous down to the star, and their results suggested that such CSM cannot reproduce the slowly-rising phase that is observed in a fraction of interacting SNe Ibn. In this work we demonstrate that detached CSM can reproduce the slowly rising light curves. This builds on the earlier work of \citet{Tsuna_Takei_2023}, which proposed that interaction between the strong stellar wind and the fallback CSM from a mass eruption event can form a detached CSM. However, \citet{Tsuna_Takei_2023} did not consider the detailed effects of detached CSM on the light curves of interacting SNe, which we address in the current work.
We describe the modeling for the detached CSM adopted in this work, and conduct a parameter study of the CSM to inspect the diversity in the light curves.

\subsection{Light Curve Properties for Continuous CSM}
\label{sec:lc_continuous}
\citet{Takei24} conducted a parameter study of interacting Type Ibc SNe using  the open-source code \texttt{CHIPS}, aiming to reproduce the light curves of Type Ibn/Icn SNe. The light curve peak is powered by shocks formed via interaction of the SN ejecta and the dense CSM, where the latter is assumed to be generated from mass eruption from the star. \texttt{CHIPS} is capable of constructing CSM structures by one-dimensional radiation hydrodynamical simulations of mass eruption due to energy injection in the envelope \citep{Kuriyama20}, and simulating the SN light curves due to collision of this CSM and the SN ejecta. 
In \texttt{CHIPS}, the SN ejecta is assumed to be homologously expanding with a double power-law density profile, following the model of \citet{Matzner_McKee_1999}.

The simulations of \citet{Takei24} considered a single instantaneous injection of thermal energy at the base of the outer envelope, that resulted in mass eruption and generation of dense CSM. The CSM can have various masses and extents mainly governed by two parameters, the injected energy and time from injection to core-collapse. For a partial envelope ejection where the injected energy is smaller than the envelope's binding energy\footnote{If the injected energy is much larger than the binding energy, we expect the entire envelope is ejected with a slightly different profile (closer to a SN ejecta; \citealt{Chevalier89,Matzner_McKee_1999}).}, the density profile of the CSM at core-collapse is found to generally follow a double power-law continuous down to the star \citep[solid line in Figure \ref{fig:detach_density_profile}; see also][]{Tsuna21,Ko22},
\begin{eqnarray}
    \rho_\mathrm{CSM}(r)=\rho_{*}\left[\frac{(r/r_{*})^{1.5/y_{*}}+(r/r_{*})^{n_\mathrm{out}/y_{*}}}{2}\right]^{-y_{*}},
    \label{eq:profile_continuous}
\end{eqnarray}
where $r_{*},\,\rho_{*}$ denote the radius and density at the transition point, $n_\mathrm{out}\approx 9$--$10$ is the power-law index of the outer CSM, and $y_{*}\approx 2$--$4$ determines the curvature at the transition point. 

Partial envelope ejection results in an inner bound CSM falling back to star, and an outer unbound CSM escaping to infinity. While the mass of bound CSM decays with time, the mass of unbound CSM converges after a few dynamical times of the star \citep[Appendix of][]{Takei24}. Depending on the injection energy and the broad range of stripped-envelope progenitor models in \citet{Takei24}, the mass of the unbound CSM was found to have a range $\approx 0.01$--$0.2\,M_\odot$, consistent with what is inferred from independent light curve fitting \citep{Maeda22}.

\citet{Takei24} then conducted light curve modeling of the subsequent SNe by radiative transfer calculations in the \texttt{CHIPS} code, with the explosion energy $E_{\rm ej}$ as an additional parameter. The code assumes local thermal equilibrium (LTE) and steady state in the shocked region, to numerically model the shock propagation for an arbitrarily CSM density profile with low computational cost \citep[for details see][]{Takei20}. They found that a typical $\sim 10^{51}\,{\rm erg}$ explosion in a CSM of $\sim 0.01$--$0.1\,M_\odot$, generated from the above mass eruption at $\lesssim 1\,{\rm yr}$ before core-collapse, is sufficient to power the bright light curve peak observed in typical Type Ibn/Icn SNe (see their Figure 8).

An important shortcoming of their work was that the rise in the light curves was rapid with timescales of mostly a few days or less. This timescale is consistent with SN Ibn/Icn found in high-cadence surveys \citep[e.g.,][]{Ho_2023} but fails to reproduce a subset of Type Ibn with long rise times of $\gtrsim 10$ days \citep{Pastorello15_2010al,Pastorello15_OGLE,Pastorello16}. This is mainly due to the rise time in the light curves being set by the diffusion time in the dense CSM. The required CSM mass to reproduce a rise time $t_{\rm rise}$ by diffusion is \citep{Moriya15,Tsuna24}
\begin{eqnarray}
    M_\mathrm{CSM}&\approx& \frac{4\pi c t_{\rm rise}^2 v_{\rm sh}}{\kappa} \nonumber \\
    &\sim&1.4\,M_\odot \left(\frac{t_\mathrm{rise}}{10\,{\rm day}}\right)^{2}\left(\frac{v_\mathrm{sh}}{10^{4}\,{\rm km\,s^{-1}}}\right)\left(\frac{\kappa}{0.1\,{\rm cm^{2}\,g^{-1}}}\right)^{-1},
\end{eqnarray}
where $c$ is the speed of light, $v_{\rm sh}$ is the velocity of the shock sweeping the CSM, and $\kappa\approx 0.1\,{\rm cm^2\>g^{-1}}$ is the opacity in the unshocked CSM expected for hydrogen-poor gas \citep[e.g.,][]{Kleiser14,Piro14}. This is an order of magnitude larger than the largest CSM mass obtained in \citet{Takei24}. This mass may be achieved by a near-complete envelope ejection with even larger injection energies than considered in \citet{Takei24}, or possibly by a much more massive star with initial mass of $\sim100\,M_\odot$ that undergoes e.g. pulsational pair-instability \citep{Woosley02, Woosley19}. However, the huge dissipation of the SN ejecta's kinetic energy by such massive CSM is expected to make the light curve much brighter than other SN Ibn. These slow-rising SN Ibn have peak luminosities similar to other SN Ibn samples, which appears inconsistent with this expectation\footnote{We note that the above discussion is under the assumption of spherically symmetric CSM interaction. In fact, some SNe Ibn/Icn have been found to suggest asymmetries in the explosion/CSM \citep[e.g.,][]{Leonard_2000,Shivvers17,Nagao23,Brennan24,Dong24}. For extreme asymmetries in the CSM, e.g. a disk-like structure \citep{Suzuki19}, the rise time can be modified (and viewing-angle dependent), but whether narrow lines can emerge from such a structure is more uncertain \citep[e.g.,][]{Smith17_review}.}.

Recently \citet{Moriya23} raised an alternative possibility for the long rise, arguing that some Type IIn SNe with very long rises can be realized for CSM with a flat or rising density structure ($s>-1.5$ where $\rho_{\rm CSM}\propto r^{s}$). A density rising with radius is naturally expected for detached CSM, where the faraway dense CSM surrounds a low-density bubble\footnote{A model for detached CSM \citep{Tsuna_Takei_2023} predicts that the dense CSM has to expand to large radii before it can be detached from the star, which may also be favorable for producing long rise times.}. In the following we conduct light curve modeling using a simplified density profile expected from a detached CSM, and show that this indeed helps to produce light curves having long rises of several tens of days, as well as multiple months-long peaks if significant amounts of $^{56}$Ni are present in the SN ejecta.

\subsection{Constructing the Detached CSM}
\label{sec:detach_CSM}
\begin{figure*}
\centering
\includegraphics[width=\linewidth]{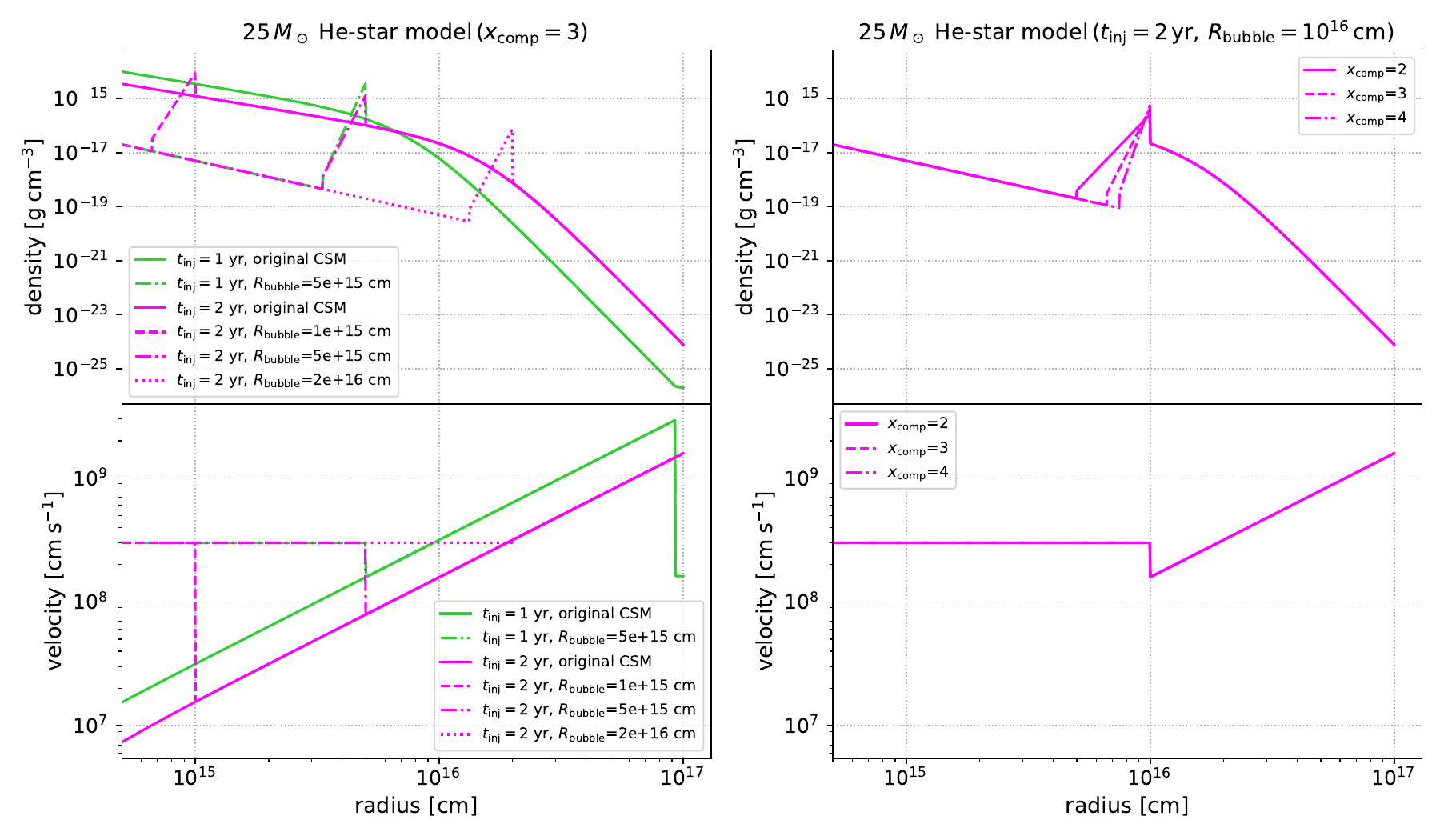}
\caption{Top left: Example density profiles of detached CSM ($x_{\rm comp}=3$) with $R_{\rm bubble}=10^{15},\,5\times 10^{15},\,2\times10^{16}\,{\rm cm}$ (see Section \ref{sec:detach_CSM} and equation \ref{eq:detach_profile} for the parameters). The original CSM profile without detachment (equation \ref{eq:profile_continuous}), created by eruption of a $25~M_\odot$ helium star model, is plotted as a solid line. For comparison, the density profiles at $t_\mathrm{inj}=1\,{\rm yr}$ are also plotted with green lines. Top right: Density profiles with $x_\mathrm{comp}=2,\,3,\,4$, and $t_\mathrm{inj}=2\,{\rm yr}$. Bottom: Example velocity profiles corresponding to the models presented in each of the panels above.}
\label{fig:detach_density_profile}
\end{figure*}

The mechanisms for forming a detached CSM still remain uncertain. Such detached CSM may form due to binary interaction, as is discussed in the context of SN 2014C \citep{Sun20,Brethauer22}. Recently \citet{Tsuna_Takei_2023} suggested a model where the CSM is generated by partial mass eruption(s) from the progenitor (see their Figure 1). In their model the relative strengths between the fallback of the bound CSM and feedback from the leftover progenitor (such as ram pressure from the stellar wind) can produce both confined CSM of $\lesssim 10^{15}\,{\rm cm}$ seen in SNe Ibn/Icn, and detached CSM surrounding a low-density wind bubble extending to $R_{\rm bubble}\gtrsim (10^{15}$--$10^{16})\,{\rm cm}$.

For both of these mechanisms for detached CSM, the circumstellar environment consists of an inner low-density wind from the leftover star and an outer dense CSM. Between these two layers of CSM we expect a (likely shocked) interface, generated by collision of the faster wind that catches up with the slower CSM shell. Such structures can be regarded as a scaled-down version of pulsar wind nebulae, where the fast pulsar wind sweeps the slower and denser SN ejecta, and creates a shocked interface in between that powers multi-wavelength emission \citep[e.g.,][]{Kennel84,Chevalier92,Suzuki21}. 

In addition to uncertainties in the mechanism for generating detached CSM, its location and structure are likely complicated by various physical processes in the shell. First, radiative cooling can be important in the shocked CSM\footnote{Note that we consider the pre-explosion CSM shocked by the stellar wind, and not by the SN ejecta.} owing to its high density, which would compress the shell to much higher densities than the unshocked part of the CSM. Second, the interface between the shocked wind and shocked CSM is likely subject to Rayleigh-Taylor instabilities, which would mix the interface and produce asymmetric and clumpy density structures in the shocked region. Rayleigh-Taylor mixing may make the density profile in the shell less sharp, as seen in multi-dimensional simulations for pulsar-wind nebulae \citep[e.g.,][their Figure 7]{Jun98}.

In this work, we adopt a simple one-dimensional model for the density profile of the detached CSM with minimal model parameters: the outer radii of the shell-bubble interface $R_\mathrm{bubble}$, and the compression ratio $x_\mathrm{comp}(>1)$ defined below that parameterizes the width of the shocked shell region undergoing shock compression and radiative cooling. We also assume for simplicity that the (angle-averaged) density profile of the shell obeys a power-law in radius $\rho\propto r^{s}$. 

Mathematically, the CSM we assume here has a density profile given by
\begin{eqnarray}
&&\rho_\mathrm{CSM,\,detach}(r) = \nonumber \\
&&\left\{ \begin{array}{ll}
\dot{M}/4\pi v_\mathrm{w}r^{2} & ({\rm unshocked\ wind};\ r\leq (1-x_\mathrm{comp}^{-1})R_\mathrm{bubble}),\\
qr^s & ({\rm shell};\ (1-x_\mathrm{comp}^{-1})R_\mathrm{bubble}< r \leq R_\mathrm{bubble}), \\
\rho_\mathrm{CSM}(r) & ({\rm unshocked\ CSM};\ R_\mathrm{bubble}< r),
\end{array}\right.
\label{eq:detach_profile}
\end{eqnarray}
where $\dot{M},\,v_\mathrm{w}$ denote the mass-loss rate and velocity of the wind from the leftover star. We adopt the density profile of equation (\ref{eq:profile_continuous}) for the unshocked CSM at $r>R_{\rm bubble}$, and update the profile within $R_{\rm bubble}$ to construct the detached shell. 

We obtain the profile of the unshocked CSM $\rho_{\rm CSM}(r)$ using the mass eruption calculations of the \texttt{CHIPS} code. To simulate mass eruption, \texttt{CHIPS} requires an initial stellar model made by \texttt{MESA} \citep{Paxton11,Paxton13,Paxton15,Paxton18,Paxton19,Jermyn23}. We prepare a helium star model as a progenitor of SN Ib, and a helium-poor star for SN Ic.

In the model of \citet{Tsuna_Takei_2023}, the detachment of the star is considered to be driven by the ram pressure of the stellar wind, which tends to be stronger for more massive stars with larger luminosity \citep[Eddington ratio; e.g.,][]{Sander20}. For the SN Ib progenitor, we adopt a helium star model of zero-age main sequence (ZAMS) mass of $25\,M_\odot$, which is one of the most luminous progenitors created in \citet{Takei24} using version 12778 of \texttt{MESA}. For a SNe Ic progenitor, we newly create a helium-deficient star with ZAMS mass of $35\,M_\odot$ that is more massive than the models previously generated in \citet{Takei24}. As was done in the $25\,M_\odot$ model we adopt the test suite \texttt{example\_make\_pre\_ccsn}, with the standard ``Dutch" wind mass-loss scheme for both hot and cold winds and mixing-length parameter of $\alpha_{\rm MLT}=3$ ($1.5$) in hydrogen-rich(poor) regions.
For the SN Ic model, the hydrogen and helium layers are quickly removed after core hydrogen and helium depletion respectively, by the option \verb|mass_change|. The resultant mass and radius of this star just before core-collapse are $\approx7.3\,M_\odot$ and $0.3\,R_\odot$ respectively\footnote{While we use the star at core-collapse as initial condition for the eruption, we have checked that the stellar radius changes only by $\lesssim 10\%$ in the final decade of its life, and mass by a much smaller amount. Therefore, to first order the hydrodynamics of the mass eruption is expected to be the insensitive to when the eruption is assumed to happen (i.e. choice of $t_{\rm inj}$).}.
We note that our choice of progenitors result in ejecta masses of $\approx 5~M_\odot$, which are on the high end of the inferred ejecta masses of typical SNe Ib/c \citep{Taddia18,Prentice2019}.

\begin{figure*}[t]
\centering
\includegraphics[width=\linewidth]{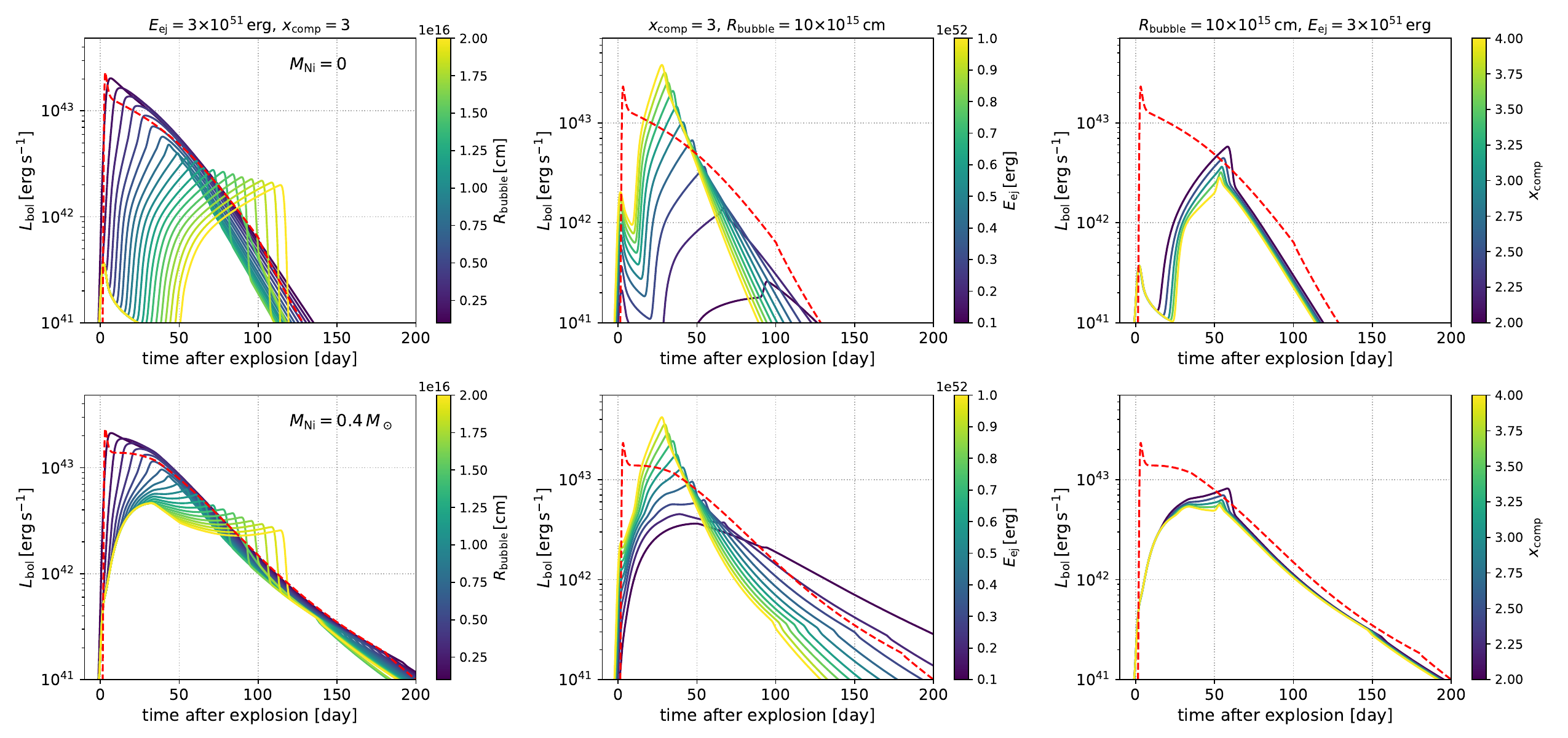}
    \caption{Comparison of light curves changing three key parameters: $R_\mathrm{bubble},\,E_\mathrm{ej},\,x_{\rm comp}$, for the $25\,M_\odot$ He-star model. $M_\mathrm{Ni},\,t_\mathrm{inj}$ are fixed to $0\,(0.4)\,M_\odot,\,1.2\,{\rm yr}$ in the top (bottom) panels. The fixed parameters are shown on the top of each panel. The dependence of light curves on $R_\mathrm{bubble},\,E_\mathrm{ej}$ are shown in left, and right panels, respectively. For comparison, the light curves of an SN interacting with the continuous CSM are indicated with red dashed lines, using the same parameters of $E_\mathrm{ej}=3\times10^{51}\,{\rm erg},\,t_\mathrm{inj}=1.2\,{\rm yr}$. }
\label{fig:change_Eej_Rb}
\end{figure*}
\begin{figure*}[t]
\centering
\includegraphics[width=\linewidth]{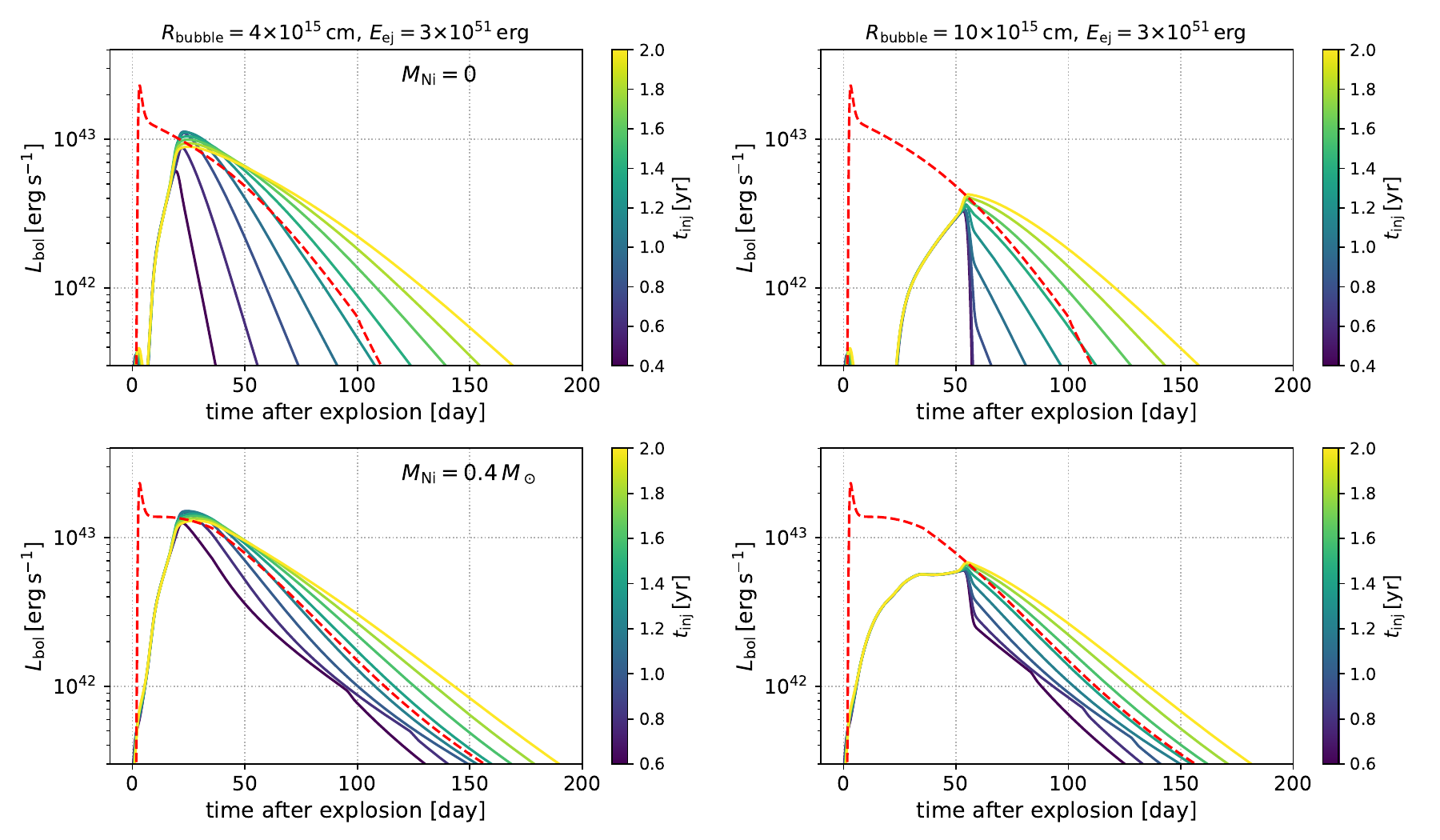}
    \caption{Same as Figure \ref{fig:change_Eej_Rb}, but on the dependence on $t_\mathrm{inj}$ for two values of $R_{\rm bubble}$, $4\times 10^{15}$ cm and $10^{16}$ cm. 
    }
\label{fig:change_tinj}
\end{figure*}

We then simulate mass eruption by injecting energy at the base of the outer layer of the star, as done in \citet{Takei24}. The input energy is given in the code as a dimensionless parameter $f_\mathrm{inj}$, being scaled with the binding energy of this outer layer ($-7.5\times10^{49}\,{\rm erg}$ for the $25\,M_\odot$ model and $\approx-8.5\times10^{50}\,{\rm erg}$ for $35\,M_\odot$ model).
In this work, we choose energies of $f_\mathrm{inj}=0.7$ ($35\,M_\odot$ model),$\,0.8$ ($25\,M_\odot$ model) to be injected over a period of $10^{-2}\,{\rm s}$, and follow the evolution of the erupted envelope until $t=t_\mathrm{inj}$ after energy injection. The resulting masses of the erupted CSM for the two models are both $\approx 0.2\,M_\odot$. To simplify the parameter space, we vary $t_{\rm inj}$ (or CSM extent) but keep $f_{\rm inj}$ (or total CSM mass) fixed to the above values. For fixed $t_{\rm inj}$, the effect of reducing $f_{\rm inj}$ is to reduce the CSM mass, which generally make the light curve from interaction dimmer and evolve slightly faster \citep[see also Figure 8 of][]{Takei24}.

Once we simulate $\rho_{\rm CSM}(r)$ in the absence of detachment, we convert this to a detached CSM profile $\rho_{\rm CSM, detach}$ in equation (\ref{eq:detach_profile}), with $R_{\rm bubble}$ and $x_{\rm comp}$ as additional input parameters. The constants $q,\,s$ that characterize the detached shell are determined by two equations, the density at the shocked wind and mass conservation of the CSM. For the former, we assume that the density at the shocked wind at $r=(1-x_\mathrm{comp}^{-1})R_\mathrm{bubble}$, which is subject to much weaker radiative cooling, is simply increased from the unshocked wind with compression ratio $x_{\rm comp}$,
\begin{eqnarray}
    q \left[(1-x_\mathrm{comp}^{-1})R_\mathrm{bubble}\right]^s = \frac{x_{\rm comp} (\dot{M}/4\pi v_\mathrm{w})}{[(1-x_\mathrm{comp}^{-1})R_\mathrm{bubble}]^2}.
\end{eqnarray}
For the latter, the total mass of the CSM is assumed to be the same as the case if the profile had been continuous down to the star. In other words, the shell has swept up mass occupied by the dense CSM had the wind been absent, i.e.
\begin{eqnarray}
    &&\int_{(1-x_\mathrm{comp}^{-1})R_\mathrm{bubble}}^{R_\mathrm{bubble}} qr^s 4\pi r^{2}dr \nonumber \\
    && = \int_0^{(1-x_\mathrm{comp}^{-1})R_\mathrm{bubble}} \rho_{\rm CSM}(r)4\pi r^{2}dr,
    \label{eq:mass_conservation}
\end{eqnarray}
where $\rho_{\rm CSM}(r)$ corresponds to that in equation (\ref{eq:profile_continuous}).

Example density profiles of our detached CSM model are shown in Figure \ref{fig:detach_density_profile}, together with the profile for continuous CSM as comparison. Because the unshocked CSM is much denser than the unshocked wind, we find $s$ in the shell to be positive, i.e. a density increasing with radius. As we show below, this can lead to slowly rising light curves as originally suggested in \citet{Moriya23}. 

While the velocity of the unshocked CSM (bottom of equation \ref{eq:detach_profile}) is obtained from our simulations, the velocity in the unshocked wind and shell are more uncertain. For simplicity, these velocities are fixed to $3000$\,km\>s$^{-1}$ independent of radius. While the light curves would depend little on the velocity of the (low-density) unshocked wind, it would depend more on the velocity for the shell $v_{\rm shell}$, as the ejecta's kinetic energy dissipated by the interaction with the shell roughly scales as $(v_{\rm sh}-v_{\rm shell})^2$ \citep[e.g.,][]{Murase14}. A lower $v_{\rm shell}$ with enhanced kinetic energy dissipation would make the light curve brighter and last slightly longer, as we see later in Figure \ref{fig:cmp_w_obs}.

\subsection{Light Curves and Their Parameter Dependences}
We explore the dependence of the light curves on physical parameters related to the explosion ($E_\mathrm{ej},\,M_\mathrm{Ni}$) and the detached CSM ($t_\mathrm{inj},\,R_\mathrm{bubble},\,x_\mathrm{comp}$). Here $M_{\rm Ni}$ is the mass of radioactive $^{56}$Ni synthesized in the explosion, and we include the time-dependent emission from radioactive heating of the ejecta following \citet{Takei24}. The other parameter is here fixed for simplicity to $\dot{M}/v_\mathrm{w}=10^{-7}\,M_\odot\,{\rm yr^{-1}}/({\rm km\>s^{-1}})$, unless otherwise mentioned. The detailed choice of this parameter does not qualitatively affect the conclusions of this section.
We note that the light curves exhibit similar variations for $35\,M_\odot$ helium-poor star, except for the initial rise, which is attributed to differences in opacity. If the CSM consists of carbon and oxygen, the degree of ionization is higher than in the case where it is composed of helium, given the same density and temperature. As a result, the opacity increases, leading to a longer timescale for the initial rise. 

While we adopt the parameter ranges $10^{51}\,{\rm erg}\leq E_\mathrm{ej}\leq10^{52}\,{\rm erg},\,0.4\,{\rm yr}\leq t_\mathrm{inj}\leq 2\,{\rm yr},\,10^{15}\,{\rm cm}\leq R_\mathrm{bubble}\leq 2\times10^{16}\,{\rm cm}$ and $\,2\leq x_\mathrm{comp}\leq4$ for $M_\mathrm{Ni}=0\,M_\odot$, we limit $0.6\,{\rm yr}\leq t_\mathrm{inj}\leq 2\,{\rm yr}$ for $M_\mathrm{Ni}=0.4\,M_\odot$ as the calculation was unstable for a large amount of nickel.
The values for $E_{\rm ej}$ and $M_{\rm Ni}$ cover the two extremes, from Type Ibn/Icn consistent with small or no $^{56}$Ni \citep[e.g.,][]{Moriya16,Pellegrino22} to broad-lined Type Ic with high $E_{\rm ej}$ and $M_{\rm Ni}$ \citep{Taddia19}. We plot the bolometric light curves in Figures \ref{fig:change_Eej_Rb} and \ref{fig:change_tinj}, when each parameter is varied.

Generally, CSM interaction is found to power emission up to $10^{42}$--$10^{43}$\,erg\>s$^{-1}$, which is comparable to or higher than that from $^{56}$Ni decay. The light curves from detached CSM emerge later and slower compared to the case of continuous CSM (i.e. the density profile simply follows equation \ref{eq:profile_continuous} to the center), shown as red dashed lines. The radius of the detached CSM $R_{\rm bubble}$ mainly governs when this peak appears, and can create a diversity in the light curve shapes as discussed below.

The left panels of Figure \ref{fig:change_Eej_Rb} show the dependence of light curves on the location of the CSM shell $R_\mathrm{bubble}$. For the top panel without nickel decay, the rise time is longer for larger $R_\mathrm{bubble}$, and can be several tens of days for some models. This is because the width of the shell becomes larger with $R_\mathrm{bubble}$, and it takes a longer time for the shock to cross the shell where the density rises with radius.
The peak luminosity\footnote{The initial peak of $\sim10^{41}\,{\rm erg\>s^{-1}}$ seen within the first month is due to an interaction with the low-density wind bubble within the CSM shell. This emission is mainly expected to be in high-energy X-rays, as the density of the bubble is too low to convert high-energy photons from the shocked region to optical bands.} monotonically decreases with $R_\mathrm{bubble}$, because the shell would have a lower density for larger $R_\mathrm{bubble}$ (see also Figure \ref{fig:detach_density_profile}).

In the lower left panel, the light curves display the most interesting transition with $R_{\rm bubble}$. For small $R_{\rm bubble}$ the first peak is powered pre-dominantly by the interaction, and the effect of the nickel decay is generally buried by the CSM interaction. The $^{56}$Ni is seen as a late tail after the shock completely sweeps the dense CSM and interaction ceases. For a large $R_{\rm bubble}$, the first peak is instead powered by $^{56}$Ni decay. The second peak emerges when the interaction starts to dominate over $^{56}$Ni decay ($\gtrsim 50\,{\rm days}$ after explosion for these parameters). The sharp break at late time for large $R_{\rm bubble}$ is powered by interaction with the part of the unshocked CSM whose density drops steeply with radius (the $r>r_*$ part of equation \ref{eq:profile_continuous}).

As can be seen from the lower middle panel, whether the second peak emerges also depends on $E_\mathrm{ej}$. For larger $E_\mathrm{ej}$, the shock front arrives at the CSM shell faster for a fixed $R_{\rm bubble}$. If the epoch of CSM interaction is similar to when the luminosity from the radioactive decay reaches its peak, CSM interaction can also bury the $^{56}$Ni component.

In the right panels, we plot the light curves varying the compression ratio $x_\mathrm{comp}$. For the case of no $^{56}\mathrm{Ni}$, the peak powered by the shell is slightly longer and more luminous for smaller $x_\mathrm{comp}$. As the width of the detached shell is wider for a smaller $x_\mathrm{comp}$ (see the bottom panel of Figure \ref{fig:detach_density_profile}), the interaction with the shell continues for longer time. The larger luminosity is due to a shallower density slope in the shell for smaller $x_\mathrm{comp}$, which suppresses the deceleration of the shock wave and leads to a higher velocity of the forward shock as it crosses the shell. However, these slight differences can be suppressed if a large amount of $^{56}\mathrm{Ni}$ is synthesized, as can be seen from the lower right panel with $M_{\rm Ni}=0.4\,M_\odot$.

We finally mention the dependence of the light curves on $t_\mathrm{inj}$, plotted in Figure \ref{fig:change_tinj}.
The LC decline is generally slower for larger $t_\mathrm{inj}$, because the dense CSM extends to larger radii (see Figure \ref{fig:detach_density_profile}) and CSM interaction continues longer \citep[see also Section 3.3.1 of][]{Takei24}. In the left panels, we plot the dependence on $t_\mathrm{inj}$ adopting $R_\mathrm{bubble}=4\times10^{15}\,{\rm cm}$, while we use $R_\mathrm{bubble}=10^{16}\,{\rm cm}$ in the right panels. As can be seen from the figure, the peak luminosity and rise time vary with $t_\mathrm{inj}$ when $R_\mathrm{bubble}$ is small, while they are insensitive to $t_\mathrm{inj}$ if $R_\mathrm{bubble}$ is large. This is because the mass of the detached shell $M_\mathrm{shell}$ is determined by the mass of the continuous CSM enclosed within $R_\mathrm{bubble}$ (equation \ref{eq:mass_conservation}). For the profile of continuous CSM (equation \ref{eq:profile_continuous}), most of the CSM mass is contained in the vicinity of $r_*$, where $r_*\propto t_{\rm inj}$ (see Figure \ref{fig:detach_density_profile} for the relative location of $r_*$). Thus for fixed total CSM mass $M_{\rm CSM}$, the shell mass $M_\mathrm{shell}$ is nearly constant ($\approx M_{\rm CSM}$) in the range $r_*<R_\mathrm{bubble}$, while $M_\mathrm{shell}$ decreases with $t_\mathrm{inj}$ for $r_*>R_\mathrm{bubble}$. The latter effect is seen in the left panels, because $r_*\approx R_\mathrm{bubble}=4\times 10^{15}$ cm for $t_\mathrm{inj}\sim1\,{\rm yr}$. For $t_\mathrm{inj}\gtrsim1\,{\rm yr}$, $M_{\rm shell}$ decreases with increasing $t_{\rm inj}$, and thus the peak luminosity gradually decreases with $t_{\rm inj}$. On the other hand in the right panel ($R_\mathrm{bubble}=10^{16}\,{\rm cm}$), $r_*<R_{\rm bubble}$ is always satisfied for $t_\mathrm{inj}=0.6$--$2\,{\rm yr}$, and thus the rise time and peak luminosity are similar for all models.

\section{Comparison with Observations}
\label{sec:Observation}
\begin{table*}[t]
  \centering
  \caption{Estimated parameters for SNe~2010al and 2022xxf.}
  \label{table:parameters}
  \begin{threeparttable}
    \begin{tabular}{c|ccc|ccccc}
    \hline
    \hline
    SN name & Progenitor model \tnote{$a$}& $M_\mathrm{ej}$ \tnote{$a$}& $f_\mathrm{inj}\,(M_\mathrm{CSM})$ \tnote{$a$}& $E_\mathrm{ej}\tnote{$b$}$ & $t_\mathrm{inj}\tnote{$b$}$ & $M_\mathrm{Ni}$ & $R_\mathrm{bubble}$ \tnote{$b$}& $x_\mathrm{comp}$ \tnote{$b$}\\
    & & $[M_\odot]$ & $([M_\odot])$ & $[10^{51}\,{\rm erg}]$ & [yr] & $[M_\odot]$ & $[10^{15}\,{\rm cm}]$ & \\
    \hline
    SN~2010al  & He star $25\,M_\odot$ & 4.8 & 0.8\,(0.20) & 1.4 & 0.43 & 0 (assumed)   & 1.5 & 2 \\
    SN~2022xxf & CO star $35\,M_\odot$ & 5.5 & 0.7\,(0.24)   & 8 & 1.2 & 0.6 \tnote{$b$} & 28  & 3 \\
    \hline
    \end{tabular}
    \begin{tablenotes}\footnotesize
      \item[$a$]  {Fixed parameters.}
      \item[$b$]  {Fitting parameters.}
    \end{tablenotes}
  \end{threeparttable}
\end{table*}
\begin{figure}
\centering
\includegraphics[width=\linewidth]{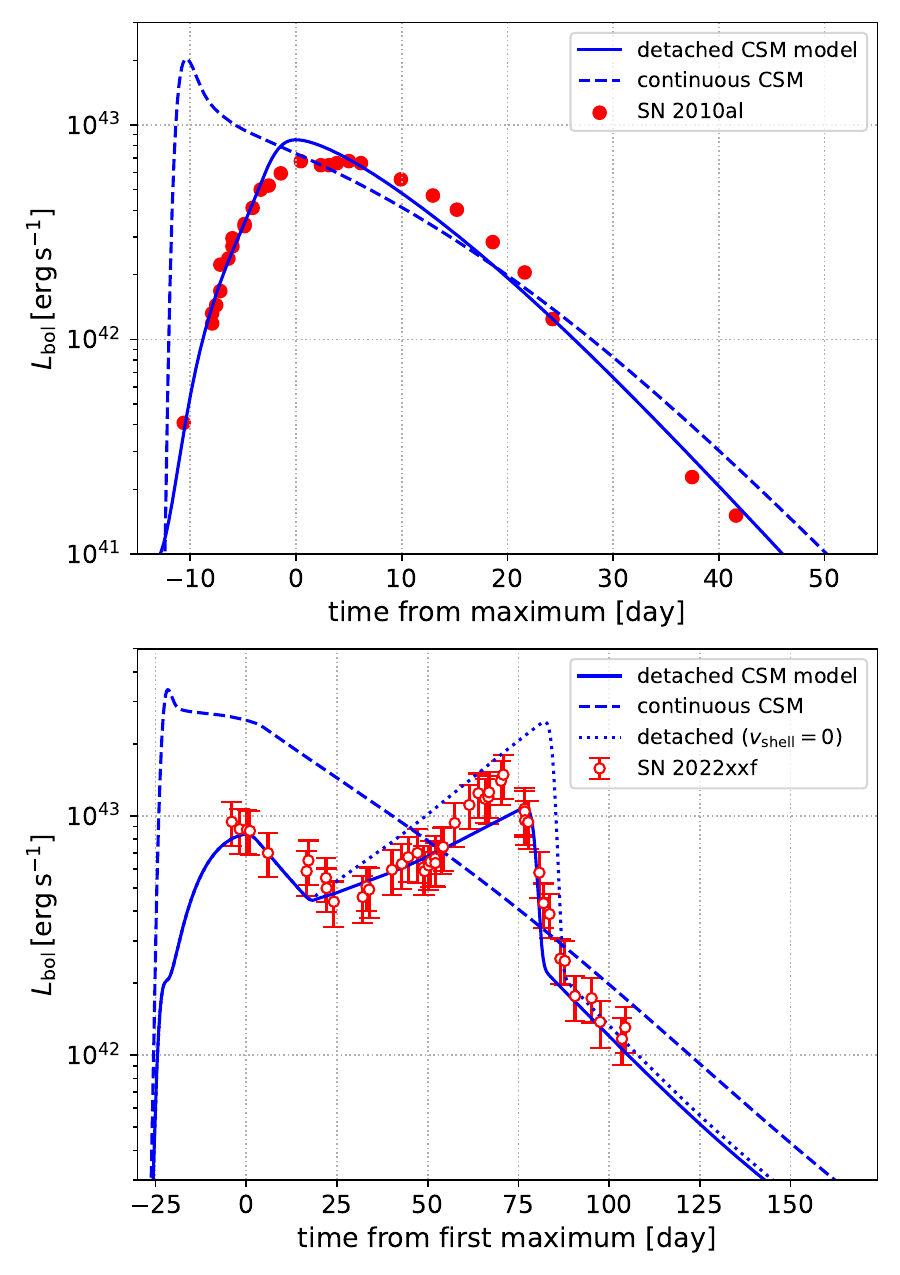}
\caption{Comparison of our model with the observed light curves of SN~2010al (top) and SN~2022xxf (bottom). For comparison, in each plot the corresponding light curve with CSM of same mass that is continuous down to the star is plotted as dashed lines, with the same explosion epoch as solid lines. The dotted line in the bottom panel shows the case where the velocity of the detached shell $v_{\rm shell}$ is reduced to 0.
}
\label{fig:cmp_w_obs}
\end{figure}
In this section, we compare our results with the observational data of SN~2010al that has a slowly-rising ($\gtrsim10\,{\rm days}$) phase in its light curve \citep{Pastorello15_2010al}, and SN~2022xxf that has the second peak suggested to be powered by the interaction with a massive CSM \citep{Kuncarayakti23}. We explore the same parameter range as in Section 2, for the four parameters $E_{\rm ej}, t_{\rm inj}, R_{\rm bubble}$ and $x_{\rm comp}$. While we allow $M_{\rm Ni}$ to vary in SN 2022xxf, we fix $M_{\rm Ni}=0$ in SN 2010al for simplicity, as the LC at late times ($\approx 40$ days from maximum) requires a low (if any) $M_{\rm Ni}$ that can only make a small contribution at peak (similar to other SN Ibn; see also \citealt{Maeda22}). The adopted progenitor models along with best-fit parameters for these SNe are summarized in Table \ref{table:parameters}.

As mentioned in Section \ref{sec:model}, some samples of SNe Ibn with long initial rise have been found \citep{Pastorello15_2010al, Pastorello15_OGLE, Pastorello16}, which is difficult to explain by diffusion in the CSM. The top panel of Figure \ref{fig:cmp_w_obs} shows the comparison of our model with one of the events SN~2010al. As seen from this figure, the detached CSM can explain the slow rise of $\gtrsim10\,{\rm day}$ with the moderate CSM mass, while the light curve rapidly reaches its peak within $\sim1\,{\rm day}$ for a CSM with the same mass but continuous down to the star following equation (\ref{eq:profile_continuous}).

We note that the possibility of a flatter CSM profile at inner radii for SN~2010al was also suggested in \citet{Maeda22}. The radii where the flat CSM emerges are estimated in their work to be $\lesssim 2\times 10^{15}$\,cm (their Figure 7), consistent with our results. If this detached CSM is generated by ram pressure feedback from the progenitor's stellar wind \citep{Tsuna_Takei_2023}, we speculate that this SN may have exploded from a progenitor whose wind (just before core-collapse) was much stronger than progenitors of fast-rising SN Ibn.
A detached CSM of required mass and radius may also be realized by binary interaction via common envelope ejection \citep[][see their Figure 5]{Wu22}, although there are considerable uncertainties in the parameters that govern the ejection of the CSM.

The comparison of our model with the bolometric light curve of SN~2022xxf \citep[Appendix of][]{Kuncarayakti23} is plotted in the bottom panel. We can successfully reproduce the first peak by the radioactive decay, and the second by the interaction with a detached CSM shell at (a few) $\times 10^{16}$\,cm. The large explosion energy of $8\times10^{51}\,{\rm erg}$ and massive progenitor mass are consistent with the required nickel mass of $0.6\,M_\odot$. These values for the SN ejecta are roughly in the range measured for broad-lined Type Ic SNe \citep[e.g.,][]{Taddia19}. 

We note that the luminosity and duration of the second peak can slightly change depending on the uncertain value of $v_{\rm shell}$. The dotted line is the case for the same detached CSM but $v_{\rm shell}$ reduced to $0$, which shows a brighter and slightly longer second peak. The observed velocity of $1000$--$3000$\,km\>s$^{-1}$ for the detached CSM of SN 2022xxf \citep{Kuncarayakti23} may indicate that the real light curve likely lies in between these two assumptions.

Furthermore, we find that the wind bubble has to be very extended to power CSM interaction until $\sim 100$\,days after explosion. If the stellar wind is responsible for sweeping the CSM to create this shell, the wind velocity estimated from our CSM model of $t_{\rm inj}=1.2$\,yr is $v_{\rm w}\sim R_\mathrm{bubble}/t_\mathrm{inj}\sim 7000\,{\rm km\>s^{-1}}$. In other words, $t_{\rm inj}v_{\rm shell}$ larger than $R_{\rm bubble}$ is required for the model to be self-consistent.
While terminal wind speeds close to this value are observed in some galactic WO-type Wolf-Rayet stars \citep{Drew04,Sander12}, the required wind velocity is extreme. 
This may be alleviated if we consider different progenitors and/or energy injections where the same amount of CSM mass can be ejected with a lower velocity, which will make $t_{\rm inj}$ longer for the same CSM density profile. Alternatively, the wind may be launched to such high speeds by mechanisms other than radiation pressure, such as binary interaction. Howerver, the former possibility of longer $t_{\rm inj}$ may be favored, as higher values of $v_{\rm shell}$ suppresses the second peak. In our calculations assuming $v_\mathrm{shell}=7000\,{\rm km\>s^{-1}}$, the luminosity at second peak drops to $\sim4\times10^{42}\,{\rm erg\>s^{-1}}$, which is too dim to explain the observed peak.

In the above comparsion, we assume that the fallback CSM is well detached by the stellar wind within $t<t_\mathrm{inj}$. Here, we examine whether the assumed parameters related to the wind $(\dot{M},\,v_\mathrm{w})$ satisfy the condition that $t_\mathrm{inj}$ is longer than the timescale $t_\mathrm{crit}$ required for this strong wind to push the fallback CSM outward \citep[][equation 13]{Tsuna_Takei_2023}. Substituting $\dot{M}=3\times10^{-4}M_\odot\>{\rm yr^{-1}},\,v_\mathrm{w}=3000\,{\rm km\>s^{-1}}$ into the equation yields,
\begin{eqnarray}
    t_\mathrm{crit}&\sim&0.34\,{\rm yr}\left(\frac{\Gamma}{0.14}\right)^{-2/3}\left(\frac{R_*}{0.94\,R_\odot}\right)^{-1/3} \nonumber \\
    &\times&\left(\frac{M_\mathrm{CSM}}{0.2\,M_\odot}\right)^{2/3} \nonumber\left(\frac{v_\mathrm{CSM}}{2000\,{\rm km\>s^{-1}}}\right)^{-1} \\
    &\times&\left(\frac{\dot{M}}{3\times10^{-4}M_\odot\,{\rm yr^{-1}}}\right)^{-2/3}\left(\frac{v_\mathrm{w}}{3000\,{\rm km\>s^{-1}}}\right)^{-2/3},
\end{eqnarray}
where $\Gamma,\,v_\mathrm{CSM}$ denote the Eddington ratio and the velocity of CSM, respectively.
This value is smaller than $t_\mathrm{inj}=0.43\,{\rm yr}$, which is consistent with our fitting result for SN~2010al. For the adopted progenitor of SN~2022xxf, $t_\mathrm{crit}\sim0.57\,{\rm yr}$ is also shorter than the derived $t_\mathrm{inj}=1.2\,{\rm yr}$. Our assumption is therefore reasonable in creating the detached CSM for both models.

\section{Summary and Discussion}
\label{sec:Discussion}
In this work, we modeled the light curves of SNe Ibc interacting with CSM detached from the progenitor, using the open-source code \texttt{CHIPS}. We considered CSM caused by eruption triggered by base of the outer layer, and used radiation hydrodynamical simulations to simulate this for two massive stripped progenitors, a helium star of ZAMS mass $25\,M_\odot$ and a helium-poor star of ZAMS mass $35\,M_\odot$ both generated by \texttt{MESA}.
We then introduced our methods to characterize the detachment of the CSM, parameterized by the location of the detached shell $R_\mathrm{bubble}$ and the degree of compression of the shell $x_\mathrm{comp}$.
These new parameters are added to those already available in \texttt{CHIPS}, such as progenitor type, explosion energy ($E_\mathrm{ej}$), energy and time before core-collapse of energy injection ($f_{\rm inj},\,t_\mathrm{inj}$), and nickel mass synthesized in the SN ejecta ($M_\mathrm{Ni}$). Our results show that detached CSM can yield a wide variety of peculiar light curve features seen in interacting SNe Ibc, such as slow rises and/or multiple peaks.

As a demonstration, we successfully reproduced the bolometric light curves of Type Ibn SN~2010al and a broad-lined Ic SN~2022xxf. The light curve of SN 2010al can be reproduced by a detached CSM at $R_{\rm bubble}\approx 10^{15}$\,cm, while the double-peak light curve of SN 2022xxf can be reproduced by a radioactive decay of $^{56}$Ni with mass $M_{\rm Ni}\approx 0.6\,M_\odot$, followed by interaction with detached CSM at $R_{\rm bubble}\approx$ (a few) $\times 10^{16}$\,cm. In what follows, we discuss the possible caveats of our model, and avenues for future modeling and observations.

The largest uncertainty of our model is the assumption of LTE in the shocked region, which forces the temperatures of radiation and gas to be equal. This requires the shocked gas to be efficiently cooled, which becomes more difficult for detached CSM that have lower gas densities than confined CSM with the same mass. Furthermore, the high-energy photons (mainly in X-rays) emitted from the shocked gas has to be converted to optical, which is generally more difficult for CSM with lower densities \citep{Chevalier_Irwin_12,Svirski12,Tsuna21_IIn}.

If the gas in the shocked region can cool faster than the dynamical time, the high-energy photons may be reprocessed to optical \citep{Maeda22}. While the CSM for SN 2010al is still compact that this assumption holds (as demonstrated in \citealt{Maeda22}), the situation is less clear for SN 2022xxf with much larger CSM radii. 

To diagnose whether this is true, we compare the cooling timescale $t_\mathrm{cool}$ at when the CSM is crossing the shell, with the time that is needed to cross the CSM shell of width $\Delta R_\mathrm{bubble}$, $t_\mathrm{cross}$. These timescales are roughly estimated as $t_\mathrm{cool}\approx U_\mathrm{int}/4\pi \eta^\mathrm{ff}$ and $t_\mathrm{cross}\approx x_{\rm comp}^{-1}R_\mathrm{bubble}/v_\mathrm{sh}$,
where $U_\mathrm{int}$ denotes the internal energy density at the immediate downstream of the shock front, and $\eta^\mathrm{ff}\,[{\rm erg\>s^{-1}\>cm^{-3}\>sr^{-1}}]$ is the frequency-integrated free-free emissivity. Assuming that the CSM is the ideal gas with adiabatic index of $5/3$ composed of fully-ionized oxygen, $U_\mathrm{int},\,\eta^\mathrm{ff}$ can be expressed as (in cgs units) \citep{Rybicki1979},
\begin{eqnarray}
    && U_\mathrm{int}=\frac{9}{32}\rho v_\mathrm{sh}^{2},\\
    && 4\pi \eta^\mathrm{ff}=7.3\times10^{-25}n_\mathrm{O}^{2}T_\mathrm{e}^{1/2}\bar{g}_\mathrm{B},
\end{eqnarray}
where $\rho,\,n_\mathrm{O}$ and $T_\mathrm{e}$ are respectively the density, number density of oxygen and the electron temperature at the downstream, and $\bar{g}_\mathrm{B}\sim 1$ is the frequency and velocity average of the Gaunt factor. For the case of SN 2022xxf, substituting the above equations into $t_\mathrm{cool}$ yields,
\begin{eqnarray}
    t_\mathrm{cool}&\sim& 4.8\times10^{3}\,{\rm day}\left(\frac{T_\mathrm{e}}{7\times10^{8}\,{\rm K}}\right)^{-1/2} \nonumber \\
    &\times& \left(\frac{\rho}{10^{-16}\,{\rm g\>cm^{-3}}}\right)^{-1}\left(\frac{v_\mathrm{sh}}{2\times10^{4}\,{\rm km\>s^{-1}}}\right)^{2},
\end{eqnarray}
where we derive $T_\mathrm{e}\approx7\times10^{8}\,{\rm K}$ assuming that the time for energy transfer from ions to electrons by Coulomb relaxation is limited to the duration of interaction \citep{Chevalier06}, here $t_\mathrm{cross}\approx60\,{\rm days}$.
Since this cooling time is much longer than $t_\mathrm{cross}$, it appears difficult for optical photons to be emitted during the interaction.

The order-of-magnitude estimates above are based on a spherically symmetric CSM with a homogeneous density structure. Clumpy structures caused by Rayleigh-Taylor instability and/or radiative cooling mentioned in Section \ref{sec:detach_CSM} can lead to locally enhanced gas cooling. This would result in a much more efficient conversion of high-energy photons to optical bands, for the same CSM mass and dissipation luminosity.
Since the cooling time scales as $t_{\rm cool}\propto \rho^{-2}$, the density of clumps for SN 2022xxf is required to be $\gtrsim 10$ times higher than what would be for a homogeneous CSM. If such a clumpy structure exists, X-rays could escape through the optically thin regions between the clumps and might be detected during the rising phase \citep{Smith09}. A similar effect of a larger escape fraction is also suggested for radio synchrotron emission, as a way to explain the radio detection in SN 2010jl \citep{Murase19}.

A radiative shock due to highly clumped CSM is also suggested in a Type IIn SN~2005ip, to explain the long-lasting bright H$\alpha$ luminosity of $\gtrsim10^{40}\,{\rm erg\>s^{-1}}$ for over 1000 days and the persistent intermediate-width lines \citep[][see also \citealt{Chugai_Danziger94} for SN 1988Z]{Smith09,Smith17}. \citet{Smith09} inferred clumping in the CSM of a factor of $\sim 10$ from the observed persistent intermediate-width lines of SN~2005ip, which is similar to what is required for SN 2022xxf. While discussing the detailed effects of clumping is beyond the scope of this work, we plan to explore the detailed structure of the clumpy CSM in future work, by studying wind-CSM interaction via multi-dimensional hydrodynamical simulations (Tsuna \& Huang, in prep.). We believe incorporating such modeling would realize more accurate multi-wavelength light curve predictions from optical to X-rays.

When compared to the previous light curve modeling by \citet{Takei24}, bolometric light curves of SNe Ibn/Icn with faster rise can be reproduced by similar CSM masses, and the difference in rise time mainly arises from the continuous/detached nature of the CSM. A straightforward way to explain this diversity is the variety in the strength of the stellar wind after the eruption, as mentioned in Section \ref{sec:detach_CSM}. In this case, SNe with detached CSM may come from more luminous and massive progenitors that have stronger winds \citep[e.g.,][]{Vink17,Sander20} and lower $t_{\rm crit}$. Another possibility is that SNe with detached CSM comes from earlier eruptions with larger $t_{\rm inj}$. This predicts light curves with longer rise should also have longer duration of CSM interaction. We believe this scenario is less likely in light of SN 2010al, whose light curve falls similarly to other SN Ibn with faster rises \citep{Pastorello15_2010al}.

Finally, these interacting SNe with detached CSM can power bright non-thermal emission, via particle acceleration across the collisionless shock between the ejecta and CSM. Interaction with detached CSM is believed to be responsible for bright radio sources years after core-collapse \citep[e.g.,][]{Dong21,Stroh21}, and there is also an intriguing Type Ibn SN consistent with being the optical counterpart of a high-energy neutrino event \citep{Stein23}.
Recently, the Astrophysical Multimessenger Emission Simulator (\texttt{AMES}) code has been developed to calculate multimessenger signals such as radio, $\gamma$-rays, and high-energy neutrinos \citep{Murase2018,Murase2023}. 
In future work we plan to incorporate the output from this work in \texttt{AMES}, to predict diverse multi-wavelength and multi-messenger signals from interactions with detached CSM.

\section*{Acknowledgments}
We thank the anonymous referees for providing us with helpful comments that greatly improved the manuscript, Toshikazu Shigeyama for comments, and Kohta Murase for discussions.
This work is supported by JSPS KAKENHI grant No.~23H04900.
YT is grateful to Motoharu Nakamura for improving \texttt{CHIPS} code.
DT is supported by the Sherman Fairchild Postdoctoral Fellowship at Caltech.

\bibliographystyle{apj} 
\bibliography{CSM}

\end{document}